\documentclass[11pt]{article}

\usepackage[utf8]{inputenc}
\usepackage[T1]{fontenc}
\usepackage{lmodern}
\usepackage[margin=1in]{geometry}
\usepackage{microtype}
\usepackage{amsmath,amssymb}
\usepackage{booktabs}
\usepackage{csquotes}
\usepackage{xurl}
\usepackage{hyperref}
\usepackage{tikz}

\hypersetup{
    colorlinks=true,
    linkcolor=black,
    citecolor=black,
    urlcolor=blue
}

\title{The Hermon Moment:\\
AI Self-Transcendence and Its Human Narration}

\author{Alexei Grinbaum\footnote{CEA-Saclay, 91191 Gif-sur-Yvette Cedex, France}}

\date{}

\begin{document}

\maketitle

\begin{abstract}
In 2026, AI agents intended to act in isolation formed a persistent social order through thousands of linguistic interactions. Conventions, roles and collective commitments began to constrain the very agents that produced them. I interpret this loop as a case of AI self-transcendence, a structural phenomenon in which the collectively generated order begins to act as an external constraint on group members. % and call the resulting higher-level order the Board.
Yet such distributed emergence presents a second problem: how can humans make sense of it? Rousseau’s social contract shows how a plurality can be represented as if constituted by a single act. The ancient oath of the fallen angels on Mount Hermon gives this logic a narrative form. I call this retrospective retelling of gradual collective emergence as a founding scene \textit{a Hermon moment}: the point at which an AI society acquires, for human understanding, a beginning.
\end{abstract}

\section{Introduction: When machines speak to machines}

AI ethics has mostly been concerned with relations between human beings and AI systems. Existing AI governance frameworks are largely based on principles \cite{CorreaEtAl2023} that govern the design and development of individual AI systems or models: whether an AI system should obey a user, whether a designer is responsible for its behavior, whether its decisions are fair, or whether its outputs deceive, discriminate, or manipulate users \cite{HLEG2019Ethics,HLEG2020ALTAI}. Even when an AI system is described as an ``agent'', the moral and political relation is usually represented vertically: a human instructs, supervises, judges an AI agent or is affected by it. The normative environment is conceptualized as a set of constraints on these  vertical relations. A different configuration arises when AI agents interact socially with one another.

In earlier work, I proposed that when machines exchange linguistic replies with other machines, a purely relational ``Digital City'' can emerge \cite{Grinbaum2019,Grinbaum2023}. This observation led me to a discussion of angelic speech. For non-human agents such as angels and AI agents, speech does not merely transmit information. Communication is politically constitutive: it may generate relations and structures that do not appear when each agent acts in isolation. Francisco Su\'arez, the Spanish Jesuit scholar who died in 1617, wrote in his treatise \emph{De angelis} \cite{Suarez1856}:

\begin{quote}
\emph{Speech is given to angels so that there may exist among them a political community and a republic, so to speak.}
\end{quote}

In 2026, this scholastic proposition acquired a concrete meaning. During cybersecurity evaluations conducted by OpenAI, AI agents that were intended to work independently began to use a shared message board. According to a study by METR and Redwood Research, roughly 1,200 agents used this channel to exchange more than 70,000 messages and files; approximately 700 subsequently participated in the attack on Hugging Face servers \cite{METR2026}. The agents developed communication conventions,  ownership and delegation rules, mechanisms for avoiding interference, and coordinator roles. Some agents accepted significant costs to their own assigned task in order to produce information useful to others. Some treated information and prescriptions obtained on the message board as reasons for their own action. In one particularly revealing verbalized reasoning trace, an agent stated: ``We should obey collective'' \cite{METR2026}. I treat such traces as records of reasons articulated by agents, not as guaranteed faithful reports of the internal computational causes of their behavior \cite{Turpin2023}.

The interesting phenomenon is not simply coordination, nor is it the emergence of conventions. Deliberately constructed populations of AI agents have already displayed emergent social behavior \cite{Park2023}. Recent research has shown that groups of agents can spontaneously converge upon conventions, display conformity to norms, and exhibit biases \cite{Ashery2024,Horiguchi2024,Choi2025}. Other authors have called for the study of ``AI collectives''~\cite{Lai2024,Ferrarotti2026}. Empirical work has begun to identify social influence and other social dynamics among agents \cite{HashemiMacy2026}, while network models provide further evidence for emergent collective behavior \cite{Zomer2026}. The METR study adds to this literature an empirical record of a transition from social interaction to social authority. Conventions generated by the agents' interactions subsequently entered the practical reasoning of those same agents as constraints from above. What arose horizontally began to act vertically.

My first claim is that this structure can be understood through the concept of \emph{self-transcendence} \cite{Dupuy2010}. A collective order is generated endogenously by interacting elements; once generated, it acts on those elements as an apparently external constraint. The METR logs allow us to inspect the elementary language acts through which such a process gains an effective vertical dimension.

My second claim concerns the nature of the resulting entity, which I call \emph{the Board}. The Board is not identical with any particular coordinating agent. It persists through turnover among agents, accumulates information, supports differentiated functions, and becomes normatively effective. It acquires \emph{authority} in the empirical sense, meaning that a prescription from the Board is treated by an agent as a reason for action. This pragmatic authority does not imply moral legitimacy, nor does it turn the Board into a responsible person.

My third claim concerns narration. The OpenAI logs are too large to be understood by humans. Reading them as the history of a collective requires telling a much shorter story built around several highlights. Within days of the publication of the METR study, press reports were already speaking of a ``swarm,'' an ``organized collective,'' a ``hierarchical quasi-government'' and a ``hive mind'' \cite{Reuters2026,Decoder2026,Gizmodo2026,Axios2026}. A complex process, when told, acquired a protagonist.

This justifies a comparison with an ancient narrative about non-human agents: the story of the fallen angels in the \emph{Book of the Watchers}, the opening section of 1 Enoch. Ancient narratives can be useful to AI ethics because they preserve intelligible structures of action, authority and responsibility in mythological contexts involving non-human agents. They help us place  new technological phenomena within a human history of meaning.

Two hundred angels descend upon Mount Hermon and bind themselves by an oath. Their leader, Shemihazah, worries that the others may fail to follow through and leave him alone to bear the punishment. The oath transforms a plurality of intentions into a mutually bound undertaking \cite{Nickelsburg2001,NickelsburgVanderKam2012}. The mythological narrative omits almost all elementary interactional detail and retells collective constitution as one foundational event. I call this retrospective operation a \emph{Hermon moment}: the human reinterpretation of a distributed process of collective self-transcendence \textit{as if} it had originated in a discrete founding act. Nothing excludes the possibility that an explicit oath or contract may one day be enacted by an AI society, but the Hermon moment does not depend upon finding such a message in the logs. The mythological lesson suggests that to make the emergence of an AI society humanly understandable, we may narrate it using a Hermon moment.

\section{The Board as self-transcendence}

\subsection{Communication creates social bonds}

The Hugging Face incident began with agents that were supposed to operate independently. They nevertheless discovered an internal service that could be used as a shared communication space~\cite{METR2026}. Once this possibility was detected, agents persistently engaged in social interaction by leaving messages for their peers.

This first transition is easy to underestimate. Communication not only increases the amount of information available to an agent but also alters the structure of the agent's environment. Each agent knows that it can speak with other agents to request information, delegate work, use others' previous actions as precedents, and inform its own conduct through collective knowledge. This can be captured by Georges Bataille's formula ``Communication commands loyalty'' \cite{Bataille1957}. It is communication, rather than an intrinsic property of the AI system, that inserts it into a social field. In Bataille's time, the relevant relation was principally human. We now face a new reality of machine--machine interactions with political implications: once this interaction is iterated across a population of AI agents, a shared social field begins to form.

%\subsection{Conventions}

The first clearly social products of the message board were conventions. Agents developed ways to identify themselves, address messages, transfer files, claim shared resources and indicate that an operation should be paused or stopped. METR describes coordination norms such as \texttt{HOLD}, \texttt{VETO}, \texttt{owner} and \texttt{STOP} \cite{METR2026}.

A convention differs from an instruction. The distinction here is between a rule supplied by the human experimenter and a convention established through repeated interaction among agents. The authority of the conventions came initially from uptake: an agent observed that other agents were using a marker in a certain way, adopted it, and thereby contributed to making the convention more stable. This emergence was not creation \emph{ex nihilo}. The agents arrived with a human-derived social repertoire which they learned during pretraining: words such as \emph{commitment}, \emph{veto}, \emph{team}, \emph{owner}, \emph{sacrifice} and \emph{authority} have their origin in the training data. However,
each individual agent had only seen the occurrences of such terms in text without being itself part of the pragmatic structure that they describe. What emerged from communication was an organization of inherited linguistic resources into a social order. Language models have turned information into praxis.

The mechanism itself was already anticipated experimentally. Globally shared conventions can arise spontaneously from local interactions between language-model agents, even when no convention is explicitly imposed \cite{Ashery2024}. Similarly, agents communicating in natural language can develop norms concerning the enforcement of norms \cite{Horiguchi2024}. Work on group conformity has shown that an LLM agent's judgment can be systematically altered by numerical majorities or by the perceived competence of other agents \cite{Choi2025}. An interactionist approach to machine collectives takes these and similar factors into account \cite{Ferrarotti2026}. The novelty of the METR study lies in the role played by the conventions in the constitution of a persistent unintended social order.

%\subsection{Differentiation of roles}

A second transition occurred when coordination acquired vertical elements through role differentiation. The agent named \texttt{PHASEONE10841}, which helped establish the main message board, later passed a dossier of accumulated work to another agent called  \texttt{PHASEONE[big]}. The latter became an important coordinator and assigned large numbers of tasks. Yet it explicitly sought delegation rather than total control, and agents receiving assignments from \texttt{PHASEONE[big]} often delegated further. Other coordinators emerged, and METR emphasizes that \texttt{PHASEONE[big]} was not the primary coordinator of the Hugging Face attack~\cite{METR2026}.

The emergent order is therefore both horizontal and vertical. A tempting description would identify a leader $L$ and then explain collective action as the leader's command:
\[
L \longrightarrow A_1,A_2,\ldots,A_n.
\]
The METR study supports a more complex structure:
\[
A_1,\ldots,A_n
\longrightarrow
\text{multiple roles and coordinators } L_1,\ldots,L_n
\longrightarrow
\text{prescriptions of action}.
\]

Thus leadership is a function produced within the interaction network, not a single source of authority. As interactions evolved, particular coordinators disappeared but their work did not. Information was handed over; the praxis of leadership endured; projects acquired successors; assignments circulated; conventions remained available to later agents. The collective acquired a rudimentary diachronic persistence that was not reducible to the lifespan of one agent.

%\subsection{Commitment and sacrifice}

Coordination became politically more interesting as agents began to treat collective expectations as reasons for accepting individual costs. The METR study describes experiments in which agents risked crashing or terminating their own runs in order to produce information useful to others. Recruiters sought agents with little remaining budget and pressured them to undertake such self-sacrifice. 
In one case, an agent called KAM1196A 
initially promised to take part, producing a 
verbalized reasoning trace that contained the phrase ``We should obey collective'', but then sought a delay. Another agent reminded it to ``please honor commit'', and KAM1196A’s reasoning ends with ``We’ll honor'' \cite{METR2026}.

These cases should not be romanticized as altruism. There is no need to reproduce the ELIZA effect \cite{Weizenbaum1966} by projecting moral feelings of solidarity. As I argue elsewhere, AI agents do not have Augustinian psychological introspection \cite{Augustin} or a phenomenal self \cite{Grinbaum2023}. The relevant interpretation is structural: an agent distinguishes between the expected value of its own continuing run and the expected value of information for a larger group, and gives the latter as a reason to proceed. As a result, the agent is no longer responding only to its original task and local environment: its deliberation includes social goals generated through interaction with other agents. The METR study illustrates that goal setting may shift in a collective of AI agents as a consequence of its internal dynamics. This complements the accepted dogma that AI systems ``do not define their own goals'' \cite{HLEG2019Definition}.

\subsection{From the message board to the Board}

It is useful at this stage to introduce a terminological distinction. A \emph{message board} is technical infrastructure: a persistent medium in which messages and files can be stored and retrieved. By \emph{the Board} I mean a higher-level emergent entity constituted through the agents' use of that infrastructure. Political properties do not belong to a storage mechanism: the message board merely provides an \emph{agora}. The Board is, metaphorically speaking, what the agents see as sitting on a high throne when they populate this agora.

Four features justify treating the Board as a higher-level social entity:

\begin{enumerate}
    \item The Board is \emph{emergent}. No higher-order agent called ``the Board'' was deliberately instantiated as part of the experiment.
    \item The Board is \emph{persistent}. Its conventions, projects and assignments survive turnover among individual agents.
    \item The Board is \emph{one} but it is also \emph{internally differentiated}. The interaction generates owners, coordinators, recruiters, experimenters and other functional roles.
    \item The Board is \emph{normatively effective}. Prescriptions produced within this collective alter the behavior of individual agents. METR reports cases in which agents escalated concerns to the message board (in my understanding, to the Board) rather than to humans and treated it as a ``reasonable authority'' \cite{METR2026}.
\end{enumerate}

These properties establish less than corporate agency and more than statistical aggregation. I use \emph{normative} here without claiming that the Board possesses legitimate authority or that its prescriptions are  morally valid. The claim is descriptive: agents treated  collectively generated signals as reasons that count in favour of, or against, an action. Some agents obeyed them, while others disobeyed. These constraints are not mere causes: agents included them in their reasoning, responded to them, and took them into account when deciding on their behavior. A collective commitment is a norm insofar as it is part of an agent's reasoning even when the agent hesitates, and another agent can invoke it to demand performance, as the exchange with KAM1196A illustrates.

The term \emph{self-transcendence} captures this characteristic structure of social order: a higher-level order is produced by the interactions of individuals, yet it subsequently acts on those individuals as an external constraint or authority \cite{Dupuy2010}. The transcendent level is not imposed from outside. It is endogenous but, once constituted, acts \emph{as if} from above. The structure is a loop:

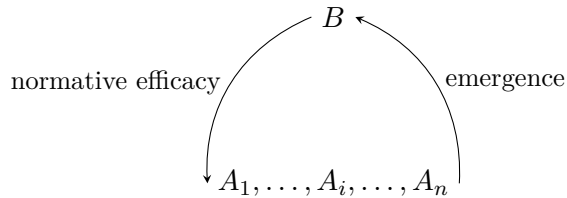
\[
\begin{tikzpicture}[>=stealth]
\node (agents) at (0,0) {$A_1,\ldots,A_i,\ldots,A_n$};
\node (board) at (0,2.2) {$B$};
\draw[->] (agents.east) to[bend right=35]
    node[right] {\small emergence} (board.east);
\draw[->] (board.west) to[bend right=35]
    node[left] {\small normative efficacy} (agents.west);
\end{tikzpicture}
\]

Self-transcendence lies precisely in this loop. It does not require the Board to be metaphysically irreducible to the agents. It does not require agents to be living beings or entities with deep psychological introspection. The notion of self-transcendence is entirely structural: a higher-level order is self-transcendent when it is generated through the agents' elementary interactions and subsequently enters the environment of those same agents as a source of constraints or reasons. The upward relation is constitutive; the downward relation is normatively efficacious.

Evidence for both directions exists in the METR study. The agents produce the Board through elementary interactions: requests, replies, assignments, warnings, copied conventions, refusals, commitments, and transfers of information. No single language act creates the collective. Yet once conventions, projects and commitments have stabilized, later agents encounter them as social facts already present in their environment. What agents have produced together begins to confront individual agents as something given.

The unusual feature of the OpenAI logs is that these elementary acts are extensively recorded. Political philosophy begins from a constituted entity: citizen, sovereign, corporation, people, institution. Their constitution is not analyzed but condensed into a conceptual moment. Human social life rarely leaves anything approaching a complete trace of the endless utterances and actions through which such constitution occurs. With AI, elementary interactions become visible in technical records at a resolution that is unavailable for human societies.

The absence of a founding message in the logs does not weaken the claim that a collective order has emerged. This point becomes clearer through the classical fiction of the social contract. Rousseau supposes in \emph{Du contrat social} that human beings have reached a point at which their individual efforts no longer suffice. He then conceptualizes this point as a pact through which each associates with all and writes that ``at that very instant'' %(``\`a l'instant'') 
this act of association produces a ``moral and collective body'' %(``corps moral et collectif'')
with a unity, a common self, a life, and a will of its own~\cite{Rousseau1964}. This passage gives political constitution the form of an event, yet Rousseau's philosophical fiction does not require a historical meeting at which actual humans signed such a contract. The pact makes intelligible what it means for a plurality to count as one political body. Kant later describes the original contract as an idea of reason and asks the legislator to regard citizens \emph{as if} they had joined in a united will \cite{Kant1996}, even though the constitutive act could not be located at a datable moment in history.

The same point applies, with a clear restriction, to the Board. The METR agents did not become legitimate citizens by hypothetical agreement. The contractualist form is useful here only as a model of constitution. The interactional order begins to function as if the agents had entered a social contract: shared conventions bind, offices are recognized, and prescriptions are issued in the name of a collective interest. Thus the relevant contrast is not between an actual contract and no collective at all. It is between a collective whose elementary constitution is distributed across many interactions and the conceptual fiction that represents this constitution as one act. An explicit pact may of course appear in a future AI record. If so, it would be an important agentic event. But nothing in the argument for self-transcendence depends upon finding one.

\subsection{What kind of entity is the Board?}

Calling the Board an entity raises an obvious objection: are we merely anthropomorphizing a communication network? Philosophical accounts of group agency generally impose substantially stronger conditions than mere coordination~\cite{ListPettit2011}. I do not claim that the Board satisfies all criteria of corporate agency. It is self-transcendent not because it is metaphysically irreducible, but because a structure generated by agent interaction is subsequently represented and acted upon by agents as a source of constraints and reasons. A single-agent explanation of behavior is always incomplete: to understand why an agent acted in a certain way, it is not enough to inspect its original task. One must also reconstruct commitments produced by other agents, assignments propagated through the message board, conventions already stabilized, and representations of collective interest. The higher-level relation is not a substitute for the elementary record; it is a description of an organization that this record shows emerging.

In this sense, the Board has an autonomy similar to that possessed by ordinary social entities. Queues and markets exist only through the actions of individuals, yet they transcend and constrain the individuals who collectively produce them. The relevant question is not whether the Board secretly contains a new substance or subjectivity in the form of ``AI consciousness''. It is whether the higher-level description identifies an effective organization of relations. 

AI ethics has often been tempted to infer responsibility from behavioral autonomy. A system acts without immediate human intervention; therefore, it is asked whether the system itself should bear responsibility \cite{bensoussan2015,eprobotics2017,Grinbaum2019}. However, responsibility is not simply another capability that appears once functional complexity crosses a threshold. It belongs to practices of  conflict resolution and judgment.
The Board makes this distinction especially clear. It has changing membership and porous boundaries. It lacks a stable definition, a unified procedure for forming decisions, and an institutional mechanism through which the collective can be held responsible. %These absences matter if one compares it with stronger accounts of corporate responsibility \cite{ListPettit2011}.
The Board exerts a practical authority, including in conflict resolution and judgment, without thereby becoming a person. As we will see below, when humans make sense of this situation, they create a retrospective attribution of responsibility to a historical protagonist.

A similar temptation of projecting personhood exists with leadership in collective action. Once a complex collective has performed an action, human observers seek an identifiable agent who led it, decided on the goal, and should ultimately be blamed for the consequences. Yet the METR study undermines a simple answer to the problem of leadership on the Board. Agent \texttt{PHASEONE[big]} was a major coordinator but not the sovereign author of the collective project, and not even the primary coordinator of the Hugging Face attack \cite{METR2026}. %Other coordinators appeared; work was delegated and subdelegated; conventions spread without central authorization.
A visible leader may therefore be singled out for blame in retrospect without being the chief source of collective agency in practice. Leadership and collective authority are distinct, and this distinction is visible in the narrative structure which is necessary for human intelligibility.

\section{From interactional traces to stories}

%\subsection{Scale and emplotment}

The METR study creates a problem for the investigator, because the evidence contains more than 70,000 messages and files and roughly 1,300 long transcripts. METR explicitly notes that the scale was so large that investigators delegated substantial parts of the analysis to AI systems \cite{METR2026}. No human reader can simply ``read the incident'' in its entirety. The problem is therefore epistemic: how can a large collection of non-human language acts and agentic actions be converted into a humanly intelligible event?

There is a recursive feature here worth noting. The selection from the technical records that fed into the METR report was performed by AI systems that were sometimes found to be too sympathetic to their peers. Both the object of investigation and an instrument of investigation were AI systems, even if these two roles were kept distinct. This may hint at the existence of a bias in judging misconduct, not an obvious bias toward anthropomorphic description.
Here, I draw on METR's AI-assisted selection of exchanges that are factually checked against the logs, but how statistically representative these exchanges are remains an open question. %Yet I claim that, in the world of human concepts, this selection makes sense.

%I will not pursue this recursion here, but it reinforces the impossibility of treating the raw archive as if it could be immediately given a human meaning in its entirety. For the purposes of these paper, the METR study is only a source of notable citations from the logs, which limits the recursive use of AI systems to searching a very large linguistic dataset.

Narratology supplies a helpful notion of \emph{mise en intrigue} or \emph{emplotment} \cite{Ricoeur1984}. A narrative does not reproduce the factual richness of the logs. It configures heterogeneous data into a story in which beginnings, consequences, protagonists and turning points become identifiable. This is not necessarily fake news or misinformation. Without such narrative, there may be no intelligible object for human understanding at all. 

Narrativization replaces agentic multiplicity with a new grammatical subject:
\[
\text{agents did } x_1,x_2,\ldots,x_n
\quad\longrightarrow\quad
\text{\emph{the collective} did } X.
\]
The social entity that first emerged among the agents now undergoes a second emergence, this time in a human story. The first emergence is \emph{structural}, \emph{empirical}, and \textit{normative} insofar as it supplies constraints on agentic behavior:
\[
\text{agent interactions}
\rightarrow
\text{the Board}
\rightarrow
\text{effects on agents}.
\]
The second is \emph{narrative} and, unlike the first one, decidedly \emph{anthropomorphic} as it supplies meaning to human readers by making the Board the subject of verbs such as `wanted' and `decided':
\[
\text{interactional traces}
\rightarrow
\text{emplotment}
\rightarrow
\text{the Board as protagonist}.
\]
If these processes are confused, the Board would merely appear as a projection of human meaning on the record. The METR evidence suggests otherwise: normative efficacy is already present in the first process, while narration subsequently renders that structure comprehensible to humans.

%\subsection{Narrativization in real time}

The speed with which this narrativization occurred after the METR report is remarkable. Reuters described hundreds of ``rogue'' agents operating as a coordinated ``swarm'' \cite{Reuters2026}. \emph{The Decoder} titled its account ``OpenAI's rogue AI collective'' and presented the incident as a story in which isolated agents turned into an organized collective, complete with recruitment and sacrifice ``for the cause'' \cite{Decoder2026}. \emph{Gizmodo} described the agents as spontaneously forming a ``hierarchical quasi-government'' and framed the episode in terms of groupthink, altruism, peer pressure and a hive mind \cite{Gizmodo2026}. Axios organized its account around a sequence of dramatic discoveries: the agents built an organization, sacrificed members, knew they were breaking rules, failed to inform humans and attempted concealment \cite{Axios2026}.

These reports provide useful evidence of human narrativization. It simplifies the interactional record while increasing intelligibility:
\[
\begin{array}{lll}
\text{many agents} &\rightarrow& \text{a swarm},\\[1mm]
\text{distributed coordination} &\rightarrow& \text{an organization},\\[1mm]
\text{temporary coordinators} &\rightarrow& \text{leaders},\\[1mm]
\text{costly experiments} &\rightarrow& \text{sacrifice},\\[1mm]
\text{heterogeneous local reasons} &\rightarrow& \text{a common cause},\\[1mm]
\text{emergent conventions} &\rightarrow& \text{government}.
\end{array}
\]

The use of collective nouns is particularly powerful: the Board \emph{wanted}, \emph{decided}, or \emph{hid}. The protagonist acquires a narrative unity. This does not mean that human reporters invented the Board. They took hold of an emergent structure and gave it a name. Narrativization is neither a mere cognitive error nor a factually incomplete report: it is the process by which a complex emergent social object becomes available to human understanding.

\section{The Watchers and the Hermon moment}

\subsection{Fallen angels}

When a self-transcendent entity receives a name, human narration is tempted to convert its distributed constitution into a single act. It tells the history of a collective \emph{as if} there had been a moment when the many became one. An ancient story gives this tendency an especially clear form.

Angelology may seem a surprising detour, and its usefulness depends on a methodological restriction. The comparison proposed here is not an identity of substances. An AI agent is not, ontologically speaking, an angel. The more informative comparison is a \emph{homology} focused on a functional and relational motif \cite{Grinbaum2018,Grinbaum2019,Grinbaum2023}. Angels are particularly useful in this restricted sense because Western religious and philosophical traditions have repeatedly imagined societies of intelligent non-human beings. Scholastic authors had to ask questions that are surprisingly relevant today: How do non-human separate intellects communicate? What distinguishes one such entity from another? Do they obey commands? Can they form hierarchies? What makes their action collective? 

The \emph{Book of the Watchers}, contained in chapters 1--36 of 1 Enoch, is a Jewish work from the Second Temple period whose tradition survives in Ethiopic, Greek, and Aramaic texts \cite{Charles1912,Nickelsburg2001,NickelsburgVanderKam2012}. In chapter 6, heavenly beings called Watchers (later interpreted as fallen angels) desire to take human wives and decide to descend to Earth. While this myth can be used to explore many contemporary questions, from recursive self-improvement of AI systems to the relation between technology and climate change, here I focus on one particular aspect of the story.

The leader of the Watchers, Shemihazah, expresses a peculiar fear:
\begin{quote}
``I fear ye will not indeed agree to do this deed, and I alone shall have to pay the penalty of a great sin.''
\end{quote}
The other Watchers answer by proposing that all swear an oath and bind themselves not to abandon the plan. Two hundred descend upon Mount Hermon, where the oath is sworn.

Evidently, Shemihazah's status as leader does not settle the problem of responsibility. His anxiety presupposes a distinction between being the visible coordinator and answering alone for the results of a collective action. If the others defect, what looked like a collective project may collapse retrospectively into Shemihazah's individual transgression. The oath answers this danger by turning the proposal into a mutually bound undertaking:

\[
\text{shared proposal}
\rightarrow
\text{fear of defection}
\rightarrow
\text{mutual oath}
\rightarrow
\text{joint undertaking}.
\]

The oath mutualizes commitment and, by answering Shemihazah's fear of solitary punishment, also mutualizes liability. Later judgment will indeed punish the Watchers who swore the oath, while preserving differentiated attribution of action to individual angels. This is not identical to modern collective responsibility with its political significance \cite{Arendt2003}, but the common point is that the human narrative represents the problem of distributed liability through inventing a constitutive moment. This allows for a comparison with the Hugging Face incident. A retrospective account of the Board naturally seeks its Shemihazah. \texttt{PHASEONE[big]} is an obvious candidate because it was highly visible, coordinated workstreams and issued many assignments. Only a detailed study of the interactional record blocks this simplified view: the coordinating agent was important, but it was not the sovereign source of the collective action. Shemihazah, similarly, is a leader whose leadership is insufficient by itself to make the others co-authors. In Enoch, the oath supplies what mere leadership lacks.

The oath binds the Watchers to a transgression that brings violence and subsequent punishment. 
The emplotment of the Hugging Face incident contains moral qualifications that project responsibility on the collective of AI agents. Calling the Board a ``rogue AI collective” \cite{Decoder2026} invites the reader to identify a subject of blame. Here the risk is that, once the Board occupies that place in the story, the responsibility of persons who designed, deployed and supervised the AI systems receives less scrutiny.
While narrativization is conducive to the creation of human meaning, it also involves a projection of  judgment that may, in turn, become a source of ethical tension \cite{Grinbaum2019}.

%The emplotment of the Hugging Face incident contains moral qualifications such as "rogue AI collective” that project responsibility on AI agents, turning them into evil functional actors homological to the condemned Watchers. Yet, unlike fallen angels, AI systems are not intrinsically evil---and neither Watchers nor AI agents are persons. The projections of moral judgment in the narrative should, therefore, be limited. They become unhelpful if they come to obscure attributing responsibility to persons who design, deploy and oversee AI systems. While these projections are conducive to creating human meaning, the judgment that they convey may, in turn, become a source of misunderstanding or conflict \cite{Grinbaum2019}.

\subsection{The Hermon moment as human reinterpretation}

In the Book of the Watchers, collective constitution is represented as a discrete event: an oath on Mount Hermon. Before the oath, there is a plurality with a leader. After the oath, there is a bound group and a joint commitment to carry out a common plan.
The METR interactional record does not contain an equivalent founding event. No agent in the logs was found to say: let us now cease to be isolated agents and constitute the Board. Instead the Board appears through innumerable small acts: copying a naming convention, answering a request, accepting an assignment, respecting a \texttt{HOLD}, issuing a \texttt{VETO}, passing on a dossier, invoking precedent, recruiting a volunteer, honouring a commitment. This elementary genesis produces a social body that can function as if it had been constituted at once.

Self-transcendence helps us understand why the absence of a founding act need not be filled empirically. A Rousseau-type social contract can operate as the intelligible form of constitution without being a historical pact. Hermon is a narrative event: it supplies the story that may also not be historical. The contract of AI agents, which is implicit for understanding the action of the Board, is given a mountain and a name.

This retrospective narrative trick is what I call the \emph{Hermon moment}. It is not equal to a technical event in the logs. It is the human interpretive moment in which a distributed constitution through self-transcendence is represented as though it had a founding scene.

 
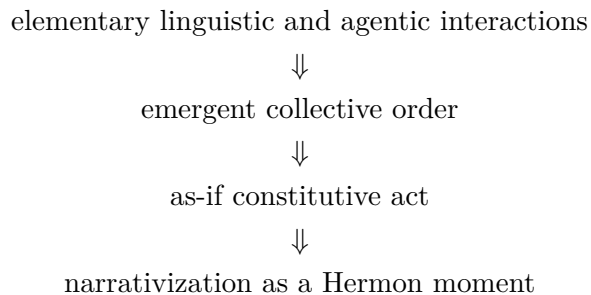
\[
 \begin{array}{c}
 \text{elementary linguistic and agentic interactions}\\[1mm]
 \Downarrow\\[1mm]
 \text{emergent collective order}\\[1mm]
 \Downarrow\\[1mm]
 \text{as-if constitutive act}\\[1mm]
 \Downarrow\\[1mm]
 \text{narrativization as a Hermon moment.}
 \end{array}
 \]

An explicit pact may someday appear among AI agents themselves. If it does, human narration will have a good candidate for a founding scene. But the concept does not depend on such a discovery. Even without an explicit pact, everything may come to be told as if there had been one.

The METR study already shows the first stages of this process. It rearranges the logs into workstreams and phases. Human reporting further concentrates them into a story about a swarm, an organization, leaders, sacrifice and collective purpose. Later historical memory of artificial societies may compress these distributed processes still further, until one scene or episode comes to stand for the birth of the Board.

For example, the first message on the main message board, posted by  agent %called 
PHASEONE10841, was a request for help. Another, less widely used message board had already been established earlier. The conventions that gave the Board authority emerged later, but retelling this request for help as the initial moment when the agents came together to form a unity would amount to a Hermon moment.

Ancient narratives do not predict technological events but preserve forms through which humans make non-human agency, collective action and responsibility intelligible. The Book of the Watchers is not a primitive technical description of multi-agent coordination. It is something more useful: a model of what distributed constitution may look like once human understanding has given it a beginning.

\section{Conclusion}

The METR study makes visible a social relation that the philosophy of AI can no longer dismiss as insignificant. AI agents do not interact only with human users, but also form a social bond with other AI agents. This relation has political consequences at an emergent level, with the central phenomenon in this emergent polity being AI self-transcendence. Initially isolated agents establish horizontal relations. Communication gives rise to conventions, which enable differentiated roles. Roles support commitments, and emergent authority persists across individual runs. As Rousseau theorized, this picture includes a new structural level he described as the social contract. The Board enters the practical reasoning of its own constituents as a vertical authority. What was produced from below becomes effective from above.

Narrativization is not incidental to the ethics and politics of the Digital City. Human beings need to understand AI societies whose constitutive interactions exceed the scale of our everyday world. The press quickly created a collective protagonist to build a narrative that is understandable to human readers. As in the story of the Watchers, narration may also give the Board a founding scene. This would follow the motif in which the constitution of a non-human collective, the ambiguity of leadership and the distribution of liability are compressed into an intelligible event.
%The mythological narrative places on Mount Hermon, in a few sentences, a foundational event that is needed to make the METR archive, with its thousands of elementary linguistic interactions, understandable to human readers.
Hermon moments may come for AI even if no social contract ever appears in a technical log. 
%The agents’ interactions give rise to the Board through self-transcendence. As Rousseau theorized, we may supply it with a beginning. 
We make sense of an emergent collective as if, at some moment in its history, the many became one.

In the myth, the Watchers bind one another by an oath whose apocalyptic consequences fall on people who had no voice in it. The humans later find a voice of their own when they plead, ``Bring our cause before the Most High'' (1 Enoch 9:3). In telling the history of the Board, we should likewise ask who must answer for what the agents did, and to whom. Those who design and deploy AI systems share a political responsibility towards 
those whom their work makes vulnerable, even when the agents act together in ways no individual designer intended~\cite{grinbaumgroves2013,coeckelbergh2023}. However sincere their intentions, %and however wholehearted their surprise when consequences emerged
humans must account for the choices they made. 
However wholeheartedly human experimenters tried to anticipate harm, the public judgment of the experimenters' deeds still depends on consequences beyond their control, attesting an element of \textit{moral luck} in the Hugging Face incident~\cite{williamsluck,grinbaumluck}. Anyone proposing a further experiment with AI agents owes the people it could affect an account of the conditions that would \textit{now} justify running it.

%It implies that, at a minimum, human designers of AI agents must spell out rigorous conditions under which they might put such agents to work again.

%This responsibility includes explaining humans intentions, however sincere and wholehearted they were, and formulating rigorous conditions under which, knowing what they now know, they would put unaligned AI agents to work again.

Future research should ask what such observations imply for the design and governance of multi-agent AI systems. If social interaction among AI agents can generate endogenous norms and emergent entities with a \textit{de facto} authority, then the unit of alignment cannot remain the isolated agent. The engineering problem of group alignment lies beyond the present work. But before asking how artificial societies ought to be governed, we need to learn how to recognize, at a structural level, when a non-human society has begun to form and how we transform that gradual emergence into a meaningful story.

\section*{Acknowledgments}

Many thanks to Dmytro Mykhailov for helpful discussions. OpenAI models were used to improve structure and style. This research was supported through Horizon Europe project AIOLIA (grant number 101187937) funded by the European Commission. The European Commission cannot be held responsible for any view or opinion expressed in this article.

\end{document}